\documentclass[twocolumn,showpacs,amsmath,amssymb,jcp,aip]{revtex4}
\usepackage{graphicx}
\usepackage{dcolumn}
\usepackage{color}
\usepackage{bm}
\usepackage{float}
\usepackage{gensymb}
\begin{document}

\title{First principles study of bimetallic Ni$_{13-n}$Ag$_n$ nano-clusters
($n =$ 0$-$13) : Structural, mixing, electronic and magnetic properties }

\author{Soumendu Datta}
\author{A. K. Raychaudhuri}
\author{Tanusri Saha-Dasgupta}
\affiliation{Thematic Unit of Excellence on Computational Materials Science, S.N. Bose National Centre for Basic Sciences, JD Block, Sector-III, Salt Lake
City, Kolkata 700 106, India}

\date{\today}

\begin{abstract}
Using spin polarized density functional theory (DFT) based calculations, combined with ab-initio molecular dynamics simulation, we carry out 
a systematic investigation of the bimetallic Ni$_{13-n}$Ag$_n$ nano clusters, for all compositions. This includes prediction of the
geometry, mixing behavior, and electronic properties. Our study reveals a tendency towards formation of a core-shell like structures, following 
the rule of putting Ni in high coordination site and Ag in low coordination site.  Our calculations predict negative mixing energies for the entire 
composition range, indicating mixing to be favored for the bimetallic small sized Ni-Ag clusters, irrespective of the compositions. The magic 
composition with the highest stability is found for the NiAg$_{12}$ alloy cluster. We investigate the microscopic origin of core-shell like structure
with negative mixing energy, in which the Ni-Ag inter-facial interaction is found to play role. We also study the magnetic properties of the 
Ni-Ag alloy clusters. The Ni dominated magnetism, consists of parallel alignment of Ni moments while the tiny moments on Ag align in anti-parallel to Ni moments. The hybridization with Ag environment causes reduction of Ni moment.
\end{abstract}

\maketitle

\section{\label{sec:intro}Introduction}
Nano systems are associated with unique physico-chemical properties because of their finite size, shape, quantum confinement as well as surface effects, and have potential applications in diverse fields from catalysis to medical diagnosis.\cite{alloy_rev1,alloy_rev2,book,another1,another2,nanodiag} During the last few decades, bimetallic nano-clusters have received special interest in the research of nano science and nano technology because of the additional degrees of freedom achievable by changing the chemical composition and ordering of its constituent atoms. Recently, several bimetallic multifunctional nano systems consisting of magnetic transition metal (M) and nonmagnetic noble metal (NM) atoms such as Fe-Pt,\cite{fept} Fe-Pd,\cite{fepd} Fe-Au,\cite{feau1,feau2} Fe-Ag,\cite{feag} Co-Pt,\cite{copt,copt2} Co-Pd,\cite{copd} Co-Au,\cite{coau} Ni-Pt,\cite{nipt} Ni-Pd,\cite{nipd} Ni-Au\cite{niau}, have been studied both experimentally as well as theoretically. Besides their catalytic properties, the M-NM bimetallic nano-systems have also great importance in the fields of fuel cells, magnetic storage as well as in spintronics applications.\cite{alloy_rev1,book,nm-fm} These M-NM nano alloys have also been  shown to possess enhanced  magneto-crystalline anisotropy constant and thermal stability of their magnetization compared to those of an equivalent pure cluster of the M elements.\cite{mae}

 In the present study, we will focus on Ni-Ag bimetallic nano clusters, which have attracted considerable research interest in recent time because of their emerging functionalities. Among the two constituents, the Ni atoms are characterized by the localized 3$d$-electrons, while the behavior of Ag atoms having a closed 4$d$-shell, are primarily governed by the {\sl itinerant} 5$s$ electron. Note that silver is of particular interest for its unique optical and electrical properties due to prominent surface plasmonic effects in the visible and ultraviolet light, which induce a  number of potential applications in sensing,\cite{sensors} surface enhanced Raman spectroscopy,\cite{serp} biomedical field\cite{bio} and so on\cite{else}. The magnetic constituent, nickel, on the other hand, has been widely used in different spintronics devices\cite{applications} and catalysis\cite{ni-cat}. Likewise, several novel applications such as sensors, electrical contacts and switches, catalysis and enhancement of photo-catalytic activity have been reported in recent time for the Ni-Ag binary systems.\cite{sensor,switch,cat2,photo} Enhanced catalytic performance of Ni-Ag bimetallic nano-systems has also been used for the reduction of nitroarenes and construction of highly efficient direct borohydride fuel cells.\cite{arene,fuel} Small Ni-Ag nanoalloys have also been predicted to exhibit high spin states and could be a promising candidate for single-molecule magnetism.\cite{highspin} However, to the best of our knowledge and contrary to expectation, in-depth electronic structure calculations correlating structure and property for the entire composition range of this interesting binary nano system, is lacking.

 An important issue in the study of alloy system, for bulk as well as nano phases, is the issue of mixing. Bulk Ni and Ag systems are reported to be immiscible for all compositions showing no tendency towards formation of solid solution.\cite{immiscible_niag} It is a pertinent question how this scenario changes at nano-scale. In case of a bimetallic nano cluster, the degree of mixing/segregation between the two species of atoms depends on a complex interplay of various parameters, such as atomic radii of the constituent species, strength of homo-nuclear {\sl versus} hetero-nuclear bonds, charge transfer, surface energies of the elements in their bulk phases and so on.\cite{alloy_rev1,factors} Depending on the mixing tendency between the two species of atoms, the atomic arrangement of a bimetallic nano system, can adopt several different patterns which can be broadly classified into three distinct types - ($i$) ordered or randomly intermixed nano-alloy, ($ii$) core-shell or onion-like multishell architectures and ($iii$) fully phase-segregated pattern, known as the {\sl janus} nanoalloys.\cite{alloy_rev1} The existing experimental as well as theoretical works on the M-NM binary nano-clusters in general, bring out an interesting common feature which suggests that when the components of the binary nano clusters have a miscibility gap in their bulk phases, a core-shell like chemical ordering is more likely to form in order to minimize the internal strain.\cite{nm-fm,mismatch,mismatch2,screening} The situation for the Ni-Ag nano-clusters is unfortunately confusing both from the point of view of experimental as well as theoretical studies, with some reports supporting core-shell like configuration following the general trend of M-NM binary clusters,\cite{agni1,agni2,agni3,agni4} while others report formation of homogeneous alloy structure.\cite{new1,homogeneous1,homogeneous2,homogeneous3,homogeneous4}For example, global optimization of the NiAg nano clusters using some semi-empirical potentials with the tight-binding second moment approximation indicated {\sl poly-icosahedral} structure of core-shell type for cluster sizes of 34 and 38 atoms.\cite{theory,agni-melting} On the other hand, the appearance of solid solution structured Ni-Ag nanoparticles, has also been discussed recently.\cite{frozen}

        In the present work, we aim on shedding light to this confusing situation through DFT based electronic structure calculations. In particular, we have studied the equilibrium structures, energetics, alloying {\sl versus} segregation behavior and magnetic properties of the Ni$_{13-n}$Ag$_n$ clusters, with $n$ = 0$-$13. We focus here on 13 atoms cluster size which is of particular interest because nano-clusters with 13 atoms have a high rate of occurrence in the time-of-flight mass spectrometry experiments for cluster production.\cite{13atoms} Such observations are indicative of their high stability. We have considered all possible compositions as we are interested in the trend of the variation for structure and electronic properties with the change of composition. The chosen size of the studied clusters allows us carrying out rigorous ab-initio calculations, which in turn leads to an in-depth understanding of the trend in the structural segregation and electronic properties. Our analysis presented here for the small Ni$_m$Ag$_n$ binary clusters indicates that mixing energy is negative at all compositions {\it i.e.} 13 atoms Ni$_m$Ag$_n$ clusters are lower in energy compared to the energy of $m$Ni$_{13}$ and $n$Ag$_{13}$ for all values of $m$ and $n$. The structural growth patterns indicate a clear trend towards core-shell like morphology with the less cohesive Ag atoms occupying the outer shell position supporting some of the experimental and theoretical studies.\cite{agni1,agni2,agni3,agni4} We find that the structural positioning of the more cohesive as well as smaller sized Ni atoms at the highest coordinated sites together with strong Ni-Ag inter-facial interaction provide the driving force for the core-shell structure. We believe that our analysis of structure-property relationship, presented here for the Ni$_{13-n}$Ag$_n$ alloyed nano-clusters will be helpful for understanding the trend in mixing of the M-NM nano alloy systems in general. We also analyzed the ground state magnetic structures for each composition, which should be useful for further study.

\begin{figure}
\rotatebox{0}{\includegraphics[height=2.0cm,keepaspectratio]{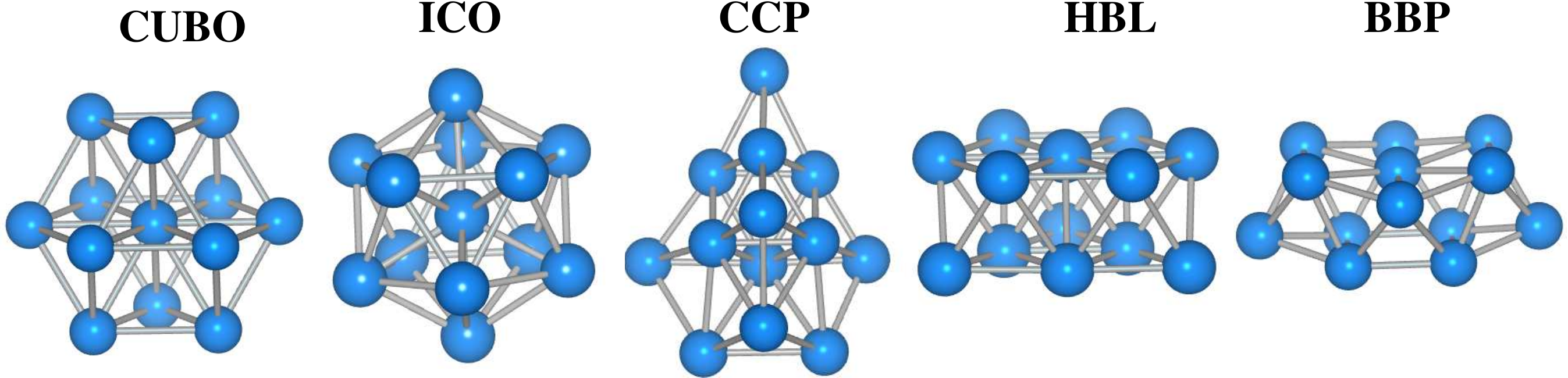}}
\caption{Starting structures used as input to optimization process within 
ab-initio molecular dynamics.}
\label{starter}
\end{figure}

\section{\label{methodology} Computational Details}
 
 The calculations reported in this study, were based on DFT within the framework of pseudo-potential plane wave method, as implemented in the Vienna ab-initio Simulation Package (VASP).\cite{kresse2} We used the Projected Augmented Wave (PAW) pseudo-potential \cite{blochl,kresse} coupled with the generalized gradient approximation (GGA) to the exchange correlation energy functional as formulated by Perdew, Burke and Ernzerhof (PBE).\cite{perdew} The 3$d$ as well as 4$s$ electrons for Ni atoms and 5$s$ as well as 4$d$ electrons for Ag atoms were treated as the valence electrons and the wave functions were expanded in the plane wave basis set with the kinetic energy cut-off of 300 eV. For the cluster calculations, a simple cubic super-cell was used with the periodic boundary conditions, where two neighboring clusters were kept separated by around 12 {\AA} vacuum space. The large cell-size essentially makes the interaction between cluster images negligible. Reciprocal space integrations were carried out at the $\Gamma$ point. The convergence of the energies with respect to the cut-off as well as k-mesh values were checked. Due to the presence of the magnetic Ni atoms, spin-polarized DFT calculations were performed. 

The task of structural optimization of a cluster, amounts to finding out the minimum energy structure (MES) on the (3$N$-6) dimensional potential energy surface (PES) of a N atoms cluster. This is a complex task for a cluster of not too small size like 13 atoms as in the present study. The situation becomes further complicated in the present case of nano-alloys due to the presence of in-equivalent permutational isomers, known as homotops.\cite{htop} This demands searching of the configurational space with a good global optimization technique. Several algorithms have been proposed for this purpose.\cite{gopt} In the present study, we have adopted the thermodynamically motivated ab-initio molecular dynamics scheme which solves the Newtonian equation of motion based on quantum mechanical forces to move from one local minima to another. To locate the global minima, we used the so-called simulated annealing approach, in which the clusters were first heated up to a temperature of T$_S$, maintained at this temperature for at least 6 ps and finally allowed to cool down again to 0 K. A time step of 1 fs was used for all calculations. The cooling process has been done for longer period and maintained at a very slow rate, with 0.05 K per iteration throughout the process. Note that such simulated annealing has been used previously for determining optimized structures of bimetallic nano-clusters.\cite{feau2,mismatch2} To speed up the global exploration of the PES, we used a favored structural motif as starting structure in the  simulated annealing process. In order to minimize the influence of starting structure, we have considered five initial morphologies, namely icosahedron (ICO), cub-octahedron (CUBO), cubic-closed packed (CCP), buckled bi-planar (BBP) and hexagonal bi-layer (HBL) structures, as shown in Fig. \ref{starter}. It should be noted that these five morphologies encompass various ordered and disordered structures reported so far for the transition metal mono-atomic clusters of 13 atoms.\cite{ni13-latest,ni13-old,ag13,mn13,fe13,co13,cubo} Importantly, these five configurations have different distribution of coordination number among the 13 constituent atoms which implies their different degrees of compactness. 

It is to be noted that the melting temperature of the Ag-rich binary (Ni$_{0.25}$Ag$_{0.75}$)$_N$ clusters with $N=$55, 147, 309 atoms, has been reported previously to be lower than 1000K.\cite{new1} However, a pure Ni$_{13}$ cluster has been predicted to be very resistance to melting with solid to liquid phase transition occurring close to the bulk melting temperature {\it i.e.} 1726 K.\cite{new2} We thus conclude while choice of T$_S$ = 1000 K would be sufficient for melting in cases of the Ag-rich clusters, the Ni-rich clusters may have melting point above 1000 K and thus the choice of T$_S$ = 1000 K may fail to escape the valley of the starting isomers. In order to ensure that, we have considered starting structures above the melting point and further repeated the simulated annealing calculations for the three Ni-rich clusters considering their structure at 2000 K temperature as the initial starting structure. Our analysis of bond length fluctuations\cite{new3,new4} during the simulated annealing runs as presented in the Supplementary Material, indicates that the melting temperature of the Ni-rich nano-alloys to be little above 1000 K. Additionally, we have considered the MES of a given composition to serve as one of the starting structures in the simulated annealing run for the other compositions.
The structures obtained out of the simulated annealing runs, have been re-optimized using the conventional zero temperature relaxation technique. In this process, symmetry unrestricted geometry optimizations were performed using the conjugate gradient and the quasi-Newtonian methods until all the force components were less than a threshold value of 0.001 eV/{\AA}. We considered both the collinear as well as non-collinear spin structures to determine the magnetic moment of the MES. While spin-orbit interaction has been considered in the non-collinear calculations, all the possible spin  multiplicities for each structure have been considered in the collinear calculations for every isomer having a given morphology for each composition. 

\begin{figure}
\rotatebox{0}{\includegraphics[height=9.6cm,keepaspectratio]{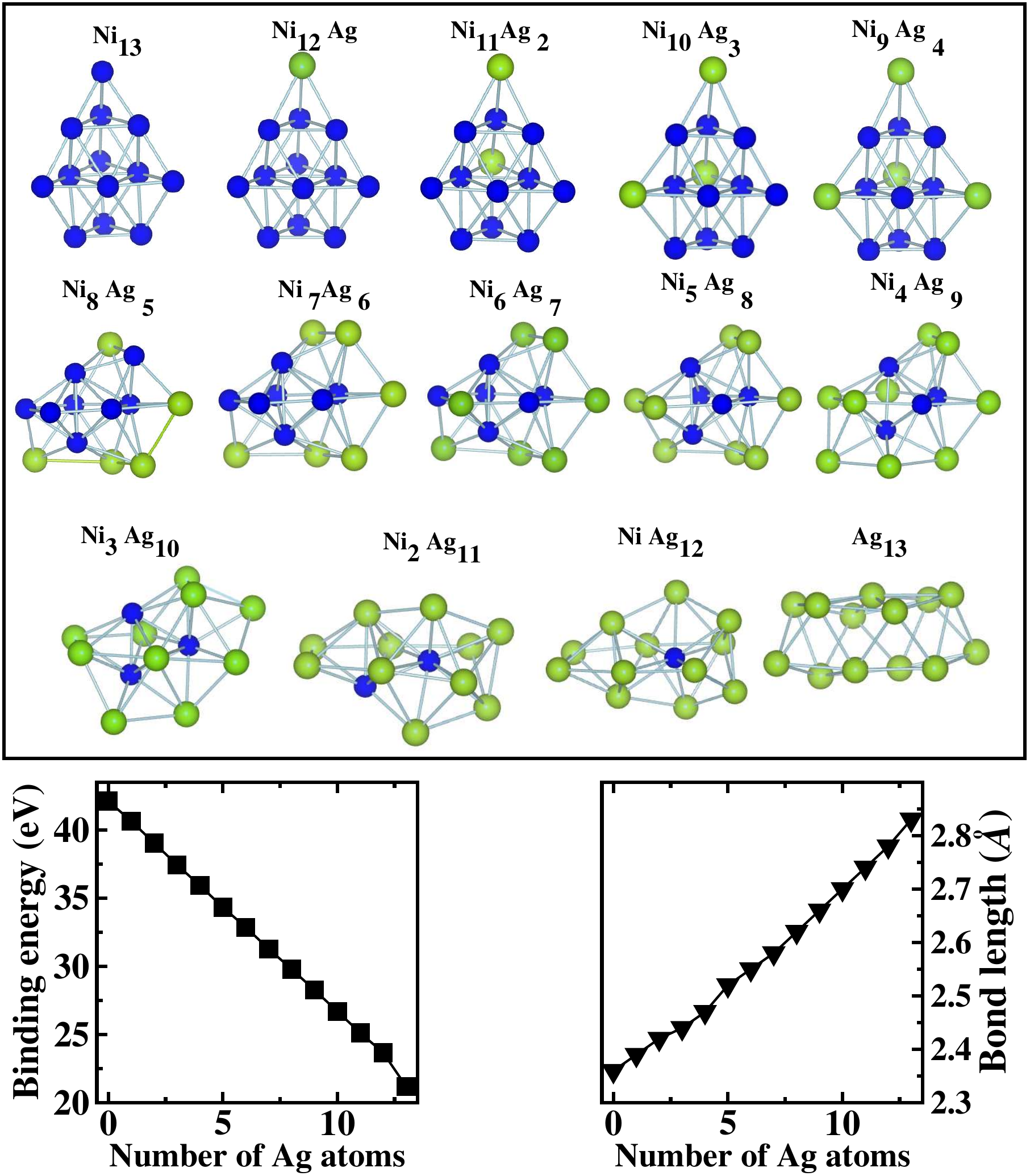}}
\caption{(Color online) Upper panel : Minimum energy structures for all Ni$_{13-n}$ Ag$_n$ clusters. The blue (darker gray) and light green (light gray) colored balls represent Ni  and Ag atoms, respectively. Bottom panels : Variations of total binding energy (left) as well as bond length (right) for the MESs of the Ni$_{13-n}$Ag$_n$ clusters.}
\label{mes}
\end{figure}

\begin{figure*}
\rotatebox{0}{\includegraphics[height=5.7cm,keepaspectratio]{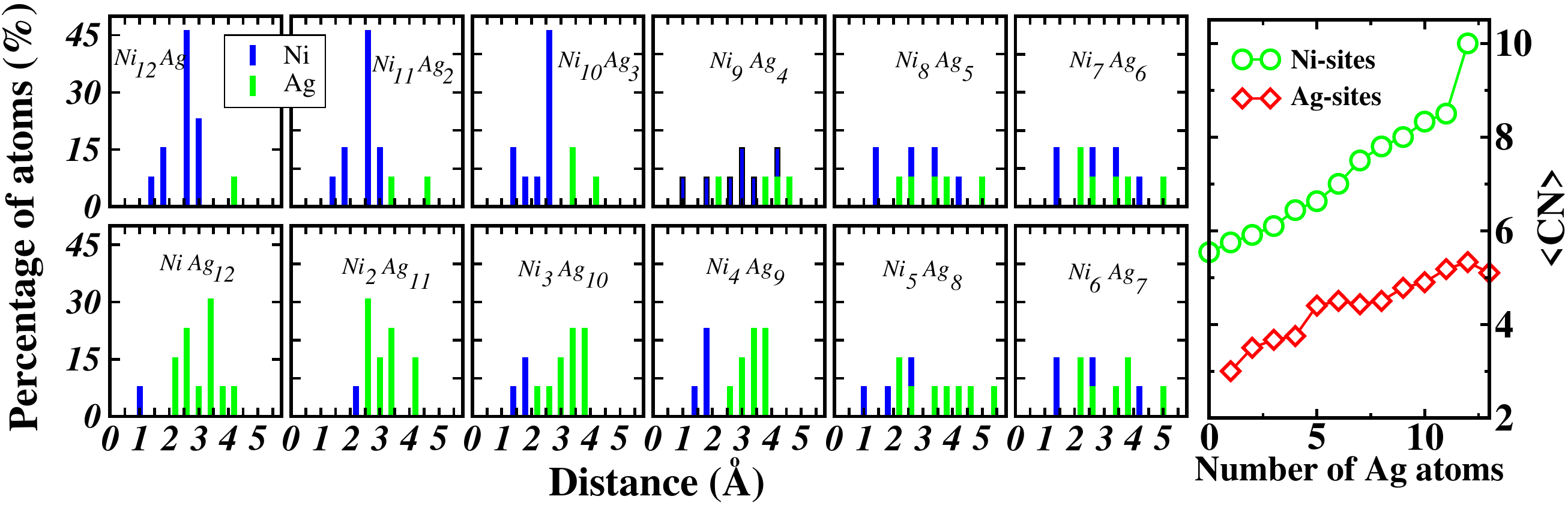}}
\caption{(Color online) Plot of radial distribution function (left panel) and average coordination number per Ni-sites as well as Ag-sites (right panel) for the MESs of the Ni$_{13-n}$Ag$_n$ clusters. The distance of the atoms in the left panel is given with respect to the center of the respective cluster. }
\label{rdf}
\end{figure*}

\section{\label{results}Results and Discussions}
In this section, we investigate the optimized ground state structures, energetics, trend in mixing, electronic and magnetic properties for the MESs of the Ni$_{13-n}$Ag$_n$ clusters.

\subsection{\label{structure}Structures }

Fig. \ref{mes} shows the minimum energy configurations of all the Ni$_{13-n}$Ag$_n$ clusters. First note that our calculations predict a CCP-like structure for the MES of the pure Ni$_{13}$ cluster and a BBP structure for the pure Ag$_{13}$ cluster in agreement with the report in literature.\cite{ni13-latest} Among the binary clusters, first four Ni-rich clusters, retain the CCP-like morphology. In these cases, the substituting Ag atoms occupy the three and four coordinated sites of the CCP structure. The use of rigorous sampling of potential energy surface of the starting structures above melting point, as explained in the Section \ref{methodology}, gives us confidence in our obtained structures. We thus believe that the preference for the CCP-symmetry based structures as the MES of the four Ni-rich binary clusters is robust and possibly guided by the special structural pattern of that morphology possessing a three-coordinated site and three four-coordinated sites. These low coordinated sites would be the obvious choice for the substituting Ag-atoms.

 For further increase of Ag content {\it i.e.} for the Ni$_8$Ag$_5$ to NiAg$_{12}$ clusters, the MESs have strongly distorted and changed their morphology completely. Interestingly, our calculations reveal a common trend in the optimized structural patterns of the Ni$_8$Ag$_5$ to NiAg$_{12}$ clusters. Each of these structures consists of a pentagonal bi-pyramidal (PBP) core of seven atoms and the rest six atoms position around this PBP core in such a way to form more pentagonal rings with the atoms of the PBP core. As for examples, the PBP core is solely made of seven Ni atoms for the MESs of the Ni$_8$Ag$_5$ and Ni$_7$Ag$_6$ clusters. Then the Ni atoms of this PBP core are getting substituted by Ag atom one by one for the subsequent Ag-rich clusters. Finally, the MES for the NiAg$_{12}$ cluster consists of several coupled PBP sections, each sharing the Ni atom either at its apex position or at the pentagonal ring.  This is a very stable configuration for the NiAg$_{12}$ cluster, where the Ni atom is coordinated by neighboring ten Ag atoms  and the Ni atom is off-centered in the cluster (cf. Fig. \ref{mes}) in order to minimize the strain arising from the size mismatch between Ni and Ag atoms. We find that an optimized ICO structure of NiAg$_{12}$ cluster with the Ni atom at the center of the ICO, is higher in energy by an amount of 0.4 eV with respect to the MES of NiAg$_{12}$ cluster. Note that using empirical potential based DFT calculations, Ferrando {\it et al.} first predicted a five-fold {\it pancake} geometry for the global total-energy minima of the 34-atoms NiAg alloy clusters.\cite{agni-melting}It is also interesting to note from the Fig. \ref{mes} that the Ni-atoms at the core part are not symmetrically positioned around the center of the binary nano-clusters. Rather, they are asymmetrically distributed off the center owing to possible strain relief, as has been suggested.\cite{ferrando}

                To analyze the energetics among the different isomers of each NiAg cluster, we have calculated their binding energy, E$_B$ which has been defined as $E_B = E_{tot}^{free-atoms}-E_{tot}^{MES}$, where E$_{tot}^{MES}$ is the total energy of the MES and E$_{tot}^{free-atoms}$ is the total energy of the constituent atoms in the gas phase ({\it i.e.} isolated form). Obviously, a more positive value of E$_B$ indicates higher binding with respect to the isolated atoms. Besides the equilibrium geometries of the two pure clusters, our calculated binding energies of 3.2 eV/atom and 1.63 eV/atom for the MESs of the Ni$_{13}$ and Ag$_{13}$ clusters respectively, are also in accordance with the recent results by the DFT calculations.\cite{ni13-latest} Our calculated binding energies for the MESs of all the NiAg clusters as shown in the left bottom panel of the Fig. \ref{mes}, decrease almost linearly with the decrease of Ni content, which is readily understandable from the trend in the bulk cohesive energies of Ni and Ag. In order to quantify the different degree of disorder in the MESs, we have calculated average effective coordination numbers ($\langle ECN \rangle$) and bond-lengths, where an exponentially averaging weight function has been employed. The weighted averaging takes into account the different contributions to the measuring quantities for the surrounding atoms at different distances in these optimized low-symmetric configurations as followed in the previous work\cite{ecn}. Among the NiAg clusters, we found that the MES of the NiAg$_{12}$ cluster has the largest value of $\langle ECN \rangle$ $\sim$5.58. Our calculated overall average bond lengths for the Ni$_{13-n}$Ag$_n$ clusters as shown in the right bottom panel of the Fig. \ref{mes}, indicate that the bond lengths increase almost linearly with the increase of Ag content. This monotonic increase of the bond lengths with the increasing Ag-content results mainly from the larger atomic radius of Ag than that of Ni.

\begin{figure}
\rotatebox{0}{\includegraphics[height=5.2cm,keepaspectratio]{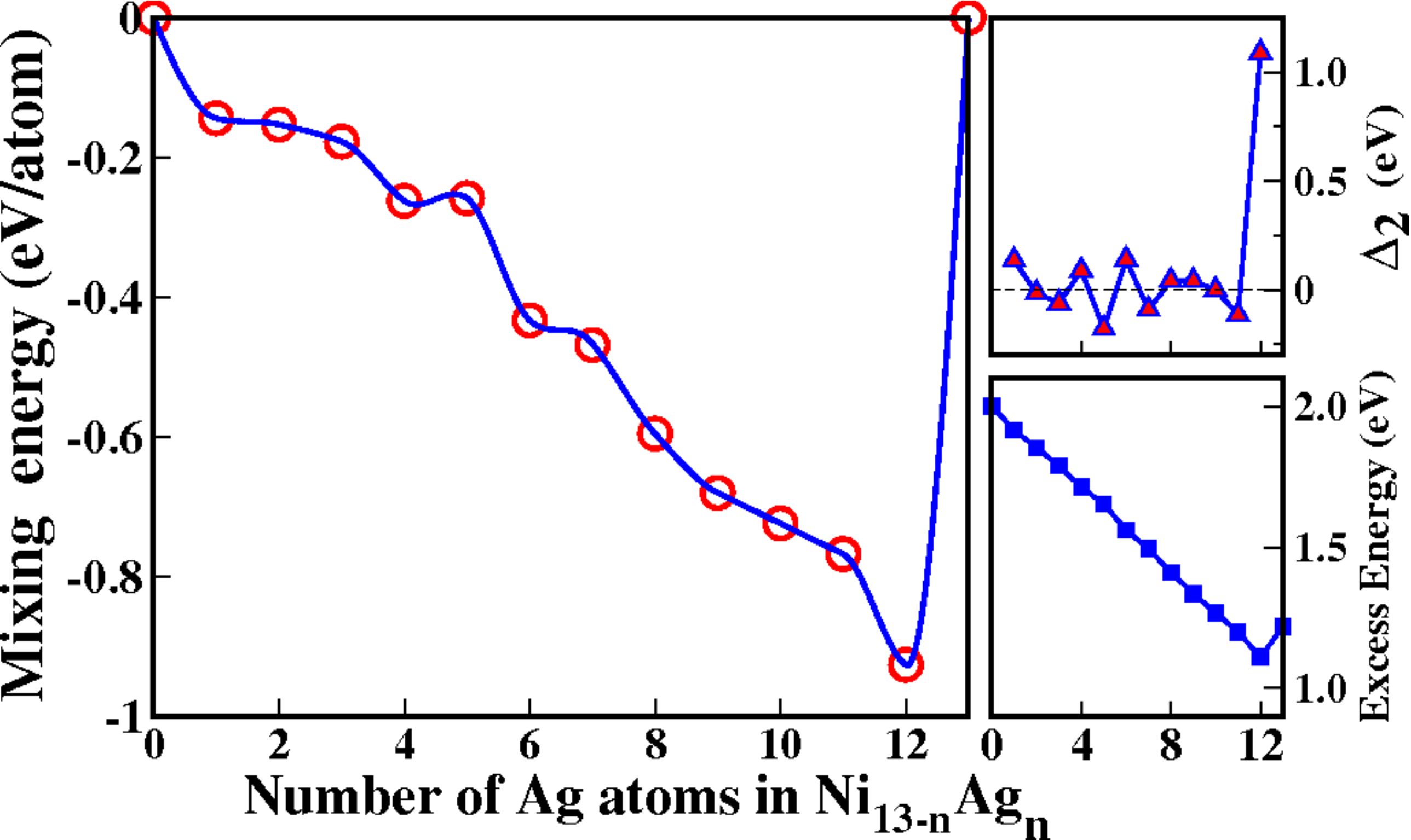}}
\caption{(Color online)Variation of mixing energies (left panel) and second differences in energy as well as excess energies (right panels) as a function of Ag-concentration for the MESs of the Ni$_{13-n}$Ag$_n$ clusters.}
\label{mixing}
\end{figure}

Regarding the chemical ordering of the two species of atoms in the MESs, it is seen from the Fig. \ref{mes} that the Ag atoms occupy mostly the less coordinated sites. On the other hand, the Ni atoms tend to occupy the more coordinated sites and they cluster together having no broken Ni$-$Ni bonds. This leads to the stability of a core-shell like structure. In order to quantify the chemical ordering of the Ni and Ag atoms in the MES of the twelve NiAg binary clusters, we have calculated the site preference of Ni and Ag atoms, measured in terms of a radial distribution function, $\eta$(r). $\eta$(r) gives the percentage of the number of Ni or Ag atoms in a spherical cell  of width $\Delta r$ at a distance between r and r$+\Delta r$ from the center of the cluster. Our calculated radial distribution function for the MESs of the twelve binary NiAg clusters, has been shown in Fig \ref{rdf}. It clearly indicates that there is a strong preference of the Ag atoms to be positioned far away from the center of the cluster, while the Ni atoms position mostly near the center of the clusters. We have also calculated the number of the nearest neighbors within a fixed cut-off distance, for every constituting atom in the MES of each Ni$_{13-n}$Ag$_n$ cluster. The variations of the average nearest neighbor coordination number per Ni-site as well as Ag-site are also shown in the Fig. \ref{rdf}. It is clearly seen that the coordination number for the Ni atoms is always higher than that of the Ag-atoms, which again support towards formation of core-shell structure with Ni atoms at the core and Ag atoms at the outer shell position, as reported earlier for the bigger sized Ni-Ag alloy clusters.\cite{theory,agni-melting,agni-coreshell}

\subsection{\label{alloy}Mixing and Electronic Properties}
To estimate the stability of the binary NiAg clusters, we have calculated mixing energy\cite{alloy_rev1,agni-coreshell,mntc} of the MESs for all compositions. The mixing energy for a 13-atoms Ni$_m$Ag$_n$ cluster ({\it i.e.} $m + n $=13) is defined as 

\begin{equation}
E_{mix}=\frac{1}{13}\left[E(Ni_mAg_n)-\frac{m}{13}E(Ni_{13})-\frac{n}{13}E(Ag_{13})\right]\nonumber
\label{mee}
\end{equation}

Here E(Ni$_m$Ag$_n$) is the energy for the MES of the binary Ni$_m$Ag$_n$ cluster, while E(Ni$_{13}$) and E(Ag$_{13}$) are the energies of MESs of the pure Ni$_{13}$ and Ag$_{13}$ clusters respectively. It gives a measure of the energy gain for the binary cluster with respect to the segregated phase consisting of two independent clusters of $\frac{m}{13}\times Ni_{13}$ and $\frac{n}{13}\times Ag_{13}$. The above equation implies that the mixing energy is always zero for the pure extremes, while its negative values indicate that the alloy formation is thermodynamically favorable. The largest negative value of the mixing energy corresponds to the most stable alloyed cluster for the given size. On the other hand, a positive value of the mixing energy will characterize a tendency of segregation among the two species of constituent atoms. Fig. \ref{mixing} shows the plot of our calculated mixing energies of all the MESs of the Ni$_{13-n}$Ag$_n$ clusters as a function of Ag atoms in them. {\sl Interestingly, the mixing energies are negative for all compositions, indicating that from the energy consideration alloy formation for the Ni$_{13-n}$Ag$_n$ clusters, is favored for the whole range of compositions.} As seen from the Fig. \ref{mixing}, the mixing energies of the MESs decrease almost monotonically with the increase of Ag-content and finally, reach the minimum value for the NiAg$_{12}$ cluster. This composition, therefore, corresponds to the {\it magic} composition having the highest stability for the 13 atoms sized NiAg alloy clusters. Importantly, most of the NiAg alloy clusters are stable by substantial energy margin, which indicates a clear propensity of mixing for the binary NiAg clusters. It is to be noted that previous study\cite{80percen} also indicates formation of the most stable NiAg alloyed nano-clusters for Ag-rich composition, in accordance with our results. The present study, however, to the best of our knowledge, reports for the first time favorable mixing of the binary NiAg nano-clusters for all possible compositions. In order to understand the trend in stability for the NiAg alloyed clusters, we have further analyzed the HOMO-LUMO gap for the MES of each composition. In accordance with our prediction of the highest stability for the NiAg$_{12}$ cluster, we find that the NiAg$_{12}$ cluster is also associated with the maximum HOMO-LUMO gap $\sim$ 0.5 eV. Combining the trend of mixing together with the trend in the structural patterns of the MESs in Fig. \ref{mes} also gives a relation between mixing and CN. It indicates that positioning of Ag atoms at the lowest coordinated sites for the Ni-rich clusters along the sequence Ni$_{12}$Ag to Ni$_7$Ag$_6$ clusters correlates with high mixing ({\it i.e.} more negative mixing energy), while positioning Ni atoms at the highest coordinated sites for the Ag-rich clusters along the sequence Ni$_6$Ag$_7$ to NiAg$_{12}$ clusters also correlates with high mixing. It implies that CN site rule is clearly playing important role for mixing.

\begin{figure}
\rotatebox{0}{\includegraphics[height=3.5cm,keepaspectratio]{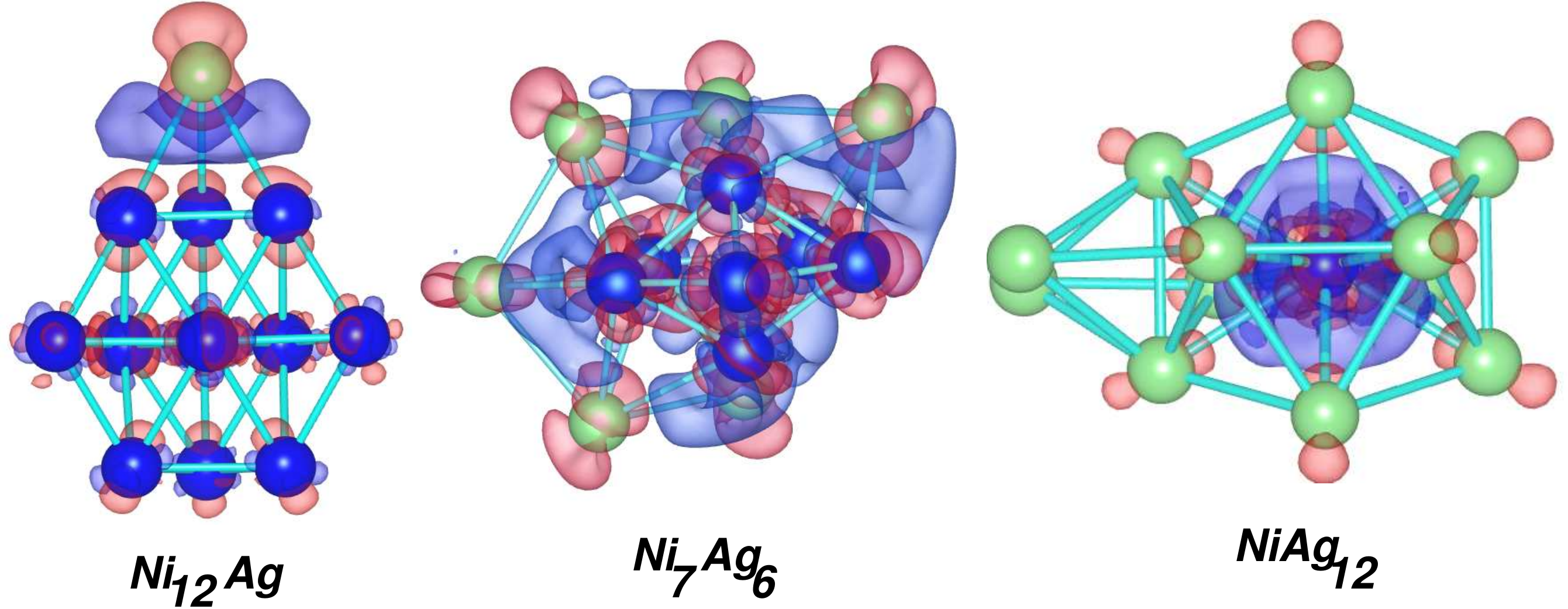}}
\caption{ (Color online) Plot of the charge density difference of the MESs for the Ni$_{12}$Ag, Ni$_7$Ag$_6$ as well as NiAg$_{12}$ clusters. The isovalue for the charge transfer plot, is fixed at 0.003 e$^{-}$/$\AA^3$. The accumulation (depletion) of charge is denoted by blue (red) colored isosurface.} 
\label{hybd}
\end{figure}

Second difference in energy, $\Delta_2$ is a quantity to compare the relative stability of a binary system between its various compositions. It is defined as\cite{alloy_rev1} $\Delta_2(13,n)= E(13,n+1)+E(13,n-1)-2E(13,n)$. The variation of $\Delta_2$ with composition for the MESs of the Ni$_{13-n}$Ag$_n$ clusters has been shown in the top right panel of the Fig. \ref{mixing}. Note that the variation of $\Delta_2$ shows the highest and the most prominent peak for $n = 12$, which reconfirms the relatively higher stability for this composition.  We have also analyzed {\sl excess energy} for the MESs of the NiAg alloyed clusters. The {\sl excess energy}\cite{agni-melting} for a binary Ni$_m$Ag$_n$ cluster with respect to $N = m + n$ bulk atoms has been defined as $E_{exc}=\frac{E(m,n)-m\epsilon_{Ni}^{coh}-n\epsilon_{Ag}^{coh}}{N^{2/3}}$; where $E(m,n)$ is the total energy of the MES for each Ni$_m$Ag$_n$ cluster and $\epsilon$'s are the bulk cohesive energies per atom. We use experimental values of the FCC bulk cohesive energies of each species, that is, $\epsilon^{coh}_{Ni}$ = -4.8332 eV/atom and $\epsilon^{coh}_{Ag}$=-2.4904 eV/atom. According to the definition, the excess energy resembles the surface energy per unit area for that cluster shape\cite{se} and the surface area is approximated by the $N^{2/3}$ factor which holds good usually for quite larger cluster size. Therefore, excess energy is a relevant parameter to compare the relative stability of the clusters of different compositions and a dip in the variation of excess energy {\sl versus} composition indicates the most stable cluster structure. The bottom right panel of the Fig. \ref{mixing} shows the plot of our calculated excess energies of the MESs for all the Ni$_m$Ag$_n$ clusters. Note that it decreases almost linearly with the increasing Ag-concentration and reaches the minimum value for the NiAg$_{12}$ cluster. Therefore, the cluster of the highest stability is not only associated with the most negative mixing energy, but also the least excess energy.

From Figs. \ref{mes} and \ref{rdf}, we learnt that a core-shell like structure is adopted for Ni$_m$Ag$_n$ clusters of all compositions, which in turn points towards the role of Ni$-$Ag interaction at the interfaces formed between its Ni-rich and Ag-rich regions. In order to investigate this, we have analyzed charge density difference $\delta\rho = \rho_{Ni_mAg_n} - \rho_{Ni_m}-\rho_{Ag_n}$; where $\rho_{Ni_mAg_n}$ is the total charge density of the Ni$_m$Ag$_n$ alloy cluster and $\rho_{Ni_m}$($\rho_{Ag_n}$) is the charge density of the Ni$_m$ (Ag$_n$) fractional cluster with same morphology as in the alloy cluster. A positive (negative) value of $\delta\rho$ would correspond to accumulated (depleted) charge. Analysis of $\delta\rho$, therefore, gives us information about the overall charge redistribution after alloying. Our calculated $\delta\rho$ values have been plotted in Fig. \ref{hybd} for the MESs of the three selected NiAg alloyed clusters, namely Ni$_{12}$Ag, Ni$_7$Ag$_6$ and NiAg$_{12}$ clusters. The plot of $\delta\rho$ indicates that there is appreciable charge transfer across the inter-atomic bonds and the accumulated charges mostly reside at the Ni-Ag interface. In case of the MES for NiAg$_{12}$ cluster, it is seen that the transferred charge is accumulated solely around the Ni atom, which also play role in destroying its magnetic character as will be discussed later.

\begin{figure*}
\rotatebox{0}{\includegraphics[height=7.9cm,keepaspectratio]{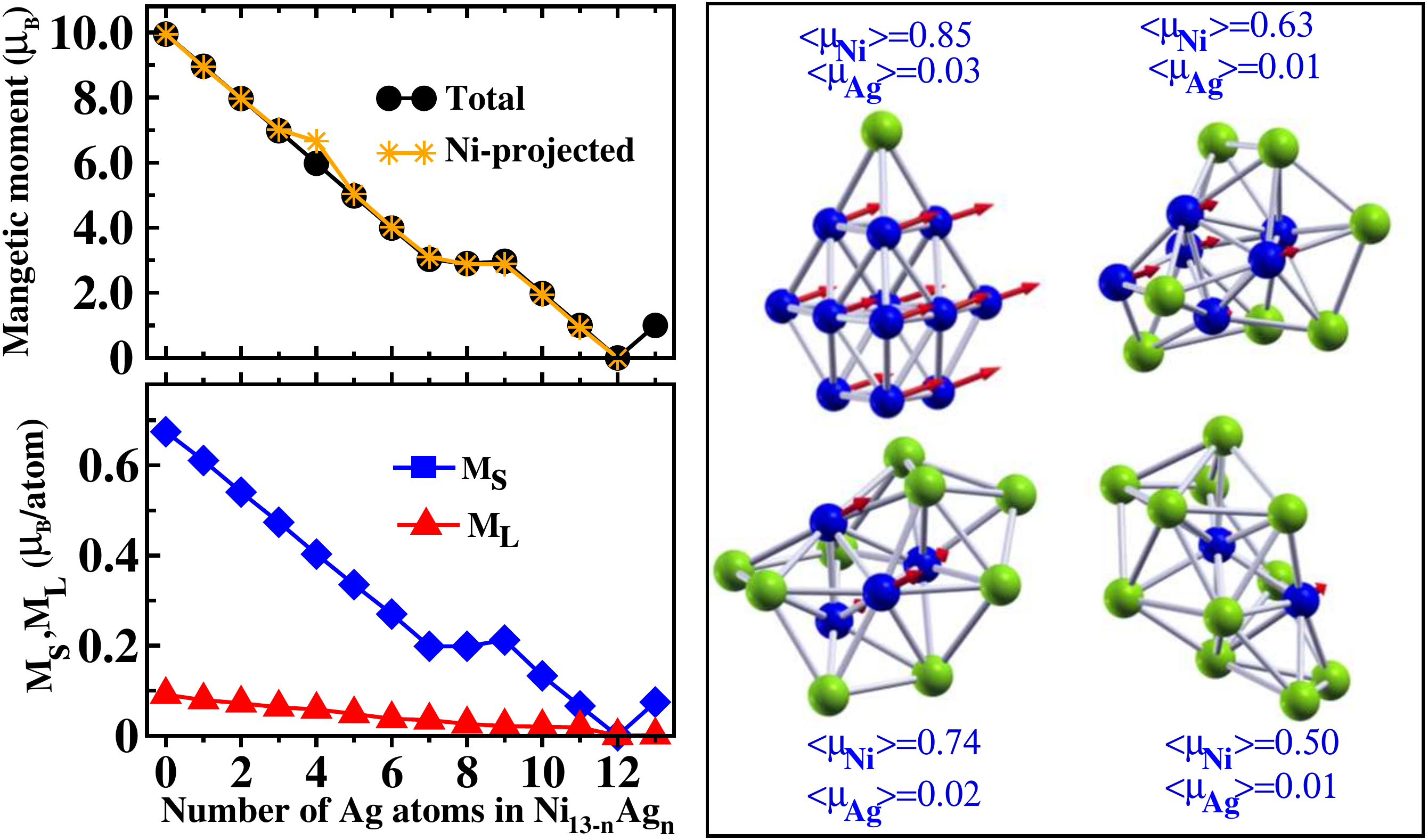}}
\caption{(Color online) Plot of overall as well as Ni projected total magnetic moments (left top panel) and average spin, orbital magnetic moments ({\it i.e.} M$_S$, M$_L$ respectively of the left bottom panel) for the MES of each NiAg system. The right panel shows the orientation and magnitude of the atom-centered magnetic moments for four alloy systems - Ni$_{12}$Ag, Ni$_6$Ag$_7$, Ni$_4$Ag$_9$ and Ni$_2$Ag$_{11}$ clusters. The values of average atom-centered magnetic moment are given in ($\mu_B$/atom). The length of the arrow is proportional to the magnitude of the atom-centered magnetic moment. Absence of arrow for the Ag atoms indicates that the magnitude of Ag-atom centered magnetic moment is very small compared to that of Ni-atom.} 
\label{magdel}
\end{figure*}

\subsection{\label{mag}Magnetic Properties}
Mono-atomic nano particles of the conventional magnetic elements usually exhibit an enhancement of magnetic moment compared to their bulk value due to the reduced coordination.\cite{fm-mag} Free-standing Ni clusters also possess magnetic moment which increases towards the atomic limit with the decreasing cluster size.\cite{ni_mag} On the other hand, nano-metric Ni clusters embedded in a metallic Ag matrix, exhibit an exceptional property. The magnetic moments of the embedded Ni clusters not only get quenched like clusters of other 3$d$ transition metal atoms,\cite{quenched} but the Ni cluster also becomes nonmagnetic below a critical size limit.\cite{nonmagnet} The calculated magnetic moments for the MESs of the binary NiAg clusters which possess a core-shell like structural pattern with Ni atoms at the core, also indicate similar type of variation and the NiAg binary cluster indeed becomes nonmagnetic in case of the NiAg$_{12}$ cluster which is incidentally identified here as the magic cluster for the 13 atoms sized binary NiAg clusters. Starting with total magnetic moment of 9.95 $\mu_B$ of pure Ni 
cluster, which is in accordance with the previous results,\cite{ni13-latest} 
the total magnetic moments of the MESs for the NiAg alloy clusters, decrease almost monotonically with the increase of Ag-content and eventually become zero for the NiAg$_{12}$ cluster, as seen from the left top panel of Fig. \ref{magdel}.
To scrutinize the contribution to the total magnetic moments, we have further analyzed the atom-centered local magnetic moment by integrating the spin density ({\it i.e.} $\rho(\vec{r})=\rho_\uparrow(\vec{r})-\rho_\downarrow(\vec{r})$) within the PAW spheres around the constituting atoms, which have radii r$_0$ = 1.5 $\AA$ for Ag atoms and r$_0$ = 1.3 $\AA$ for Ni atoms. The magnetic moment arises mainly from the localized $d$-electrons of the Ni atoms. In addition, the hybridization of the constituent Ag atoms with the neighboring Ni atoms induces small spin polarization to the Ag atoms, arising from its unpaired $s$-electron. However, the induced magnetic moment of the Ag atoms is found to be very small compared to that of the Ni atoms. Thus, as expected, the total magnetic moments show an overall decreasing trend with the decreasing Ni-atoms. 

It is important to note that though the total magnetic moment of the NiAg alloy clusters is mainly contributed by its constituent Ni atoms, its values are relatively lower compared to the magnetic moment of the free-standing pure Ni-clusters having the size same as that of the number of Ni atoms present in the corresponding NiAg alloy cluster. For example, the total magnetic moments of the isolated pure Ni$_2$, Ni$_3$ , Ni$_4$, Ni$_5$ and Ni$_6$ clusters have been predicted previously as 2 $\mu_B$, 2 $\mu_B$, 4 $\mu_B$, 6 $\mu_B$ and 8 $\mu_B$ respectively.\cite{pure_ni1,pure_ni2,pure_ni3,ni_mag} On the other hand, the total magnetic moments of the Ag$_{11}$Ni$_2$, Ag$_{10}$Ni$_3$, Ag$_9$Ni$_4$, Ag$_8$Ni$_5$ and Ag$_7$Ni$_6$ clusters in the present study, are 1 $\mu_B$, 2 $\mu_B$, 3 $\mu_B$, 4 $\mu_B$ and 5 $\mu_B$ respectively. This reduction in case of the NiAg alloy clusters, is due to the appreciable hybridization between the electronic states of Ni and Ag. The zero magnetic moment for the NiAg$_{12}$ cluster is then understandable, as the only Ni atom, in this case is coordinated by ten nearest neighboring Ag atoms and the substantial hybridization effect drives the observed
nonmagnetic state. This observation is also in line with the fact that a magnetic impurity dissolved in a nonmagnetic matrix, can indeed become nonmagnetic.\cite{ascroft} Fig. \ref{magdel} also shows separately the contributions of the spin as well as orbital magnetic moments to the total magnetic moments for the MESs of the Ni$_{13-n}$Ag$_n$ clusters. Note that there is some significant contribution of the orbital magnetic moment to the total magnetic moment. The orbital moment is contributed mainly by the Ni atoms, as seen previously for pure Ni clusters.\cite{orb_mag} The orbital magnetic moment of the NiAg alloy clusters decreases with the decrease of Ni-content, while the spin magnetic moment follows the same trend of variation for the total magnetic moment.  In the MES of the Ni$_{13-n}$Ag$_n$ clusters, it is found that the Ni-atoms centered magnetic moments are always ferromagnetically coupled irrespective of the change of composition as shown in the Fig. \ref{magdel} for some selected NiAg alloy clusters. On the other hand, the tiny Ag-centered magnetic moments (not visible in the plots of Fig. \ref{magdel}) are found to be mostly anti-ferromagnetic with the Ni-centered magnetic moments.

\section{\label{conclu}Summary and Conclusions}
In summary, we have performed first principles based electronic structure calculations for the magnetic/noble metal binary Ni$_{13-n}$Ag$_n$ clusters, which confirms that the mixing is energetically favorable for the binary NiAg clusters of all compositions. The optimal morphologies of the alloy clusters adopt a core-shell like structural patterns with the Ni-atoms positioned at the inner core region, as quantified here in terms of a radial distribution function. The magic composition, characterized with the highest mixing energy, is found for the NiAg$_{12}$ cluster, which adopts a very unusual structural pattern associated with the highest ECN, zero net magnetic moment, largest HOMO-LUMO gap and asymmetric positioning of Ni atom which is off the center of the Ag$_{12}$-cage. The Ni$-$Ag inter-facial interaction plays crucial role in determining structural, electronic and mixing properties of the alloy clusters. The zero net magnetic moment of the NiAg$_{12}$ cluster is in line with the magnetic properties of a Ni nano-cluster embedded in a metallic Ag matrix.

\section*{Supplementary Material}
See supplementary material for analysis of melting behaviour for the Ni-rich clusters in terms of root-mean-square bondlength fluctuation.

\acknowledgments
 S.D. thanks Department of Science and Technology, India for support through INSPIRE Faculty Fellowship, Grant No. IFA12-PH-27. S.D. also acknowledges 
helpful discussions with Prof. Julius Jellinek.


\begin{thebibliography}{99}
\bibitem{htop}Isomers having the same number of atoms, composition and geometric structure, but differing in the arrangement of the two species of atoms, are called homotops.
\end{thebibliography}
\end{document}